# Pressure Tuning the Mixture of $Eu^{2+}$ and $Eu^{3+}$ in $Eu_4Bi_6Se_{13}$

*Mingyu Xu*[1#], *Jose L. Gonzalez Jimenez*[1#], *Greeshma C. Jose*[2], *Artittaya Boonkird*[3,4], *Chengkun Xing*[5], *Chelsea Harrod*[6], *Xinle Li*[6], *Haidong Zhou*[5], *Alyssa Gaiser*[1,7], *Xianglin Ke*[8], *Wenli Bi*[2], *Mingda Li*[3,4], *Weiwei Xie*[1*]

1. Department of Chemistry, Michigan State University, East Lansing, MI, 48864, USA
2. Department of Physics, University of Alabama, Birmingham, AL 35294, USA
3. Quantum Measurement Group, Massachusetts Institute of Technology, Cambridge, MA, 02139, USA
4. Department of Nuclear Science and Engineering, Massachusetts Institute of Technology, Cambridge, MA, 02139, USA
5. Department of Physics, University of Tennessee, Knoxville, TN, 37996, USA
6. Department of Chemistry, Clark Atlanta University, Atlanta, GA 30314, USA
7. Facility for Rare Isotope Beams, Michigan State University, East Lansing, MI, 48864, USA
8. Department of Physics and Astronomy, Michigan State University, East Lansing, MI, 48864, USA

# equally contributed

## *Abstract*

The investigation of crystallographic, electronic, and magnetic characteristics, especially the mixed valences of $Eu^{2+}$ and $Eu^{3+}$ under pressure of a novel europium-based bismuth selenide compound, $Eu_4Bi_6Se_{13}$, presented. This new compound adopts a monoclinic crystal structure classified under the $P2_1/m$ space group (#11). It exhibits distinctive structural features, including substantial Eu-Se coordination numbers, Bi-Se ladders, and linear chains of Eu atoms that propagate along the *b*-axis. Electronic resistivity assessments indicate that $Eu_4Bi_6Se_{13}$ exhibits weak metallic behaviors. Magnetic characterization reveals uniaxial magnetic anisotropy, with a notable spin transition at approximately 1.2 T when the magnetic field is oriented along the *b*-axis. This behavior, coupled with the specific Eu-Eu interatomic distances and the magnetic saturation observed at low fields, supports the identification of metamagnetic properties attributable to the flipping of europium spins. The Curie-Weiss analysis of the magnetic susceptibility measured both perpendicular and parallel to the *b*-axis and high-pressure partial fluorescence yield (PFY) results detected by X-ray absorption spectroscopy (XAS) reveal the tendency of the material to enter a mixed valent state where the trivalent state becomes more prominent with the pressure increase or temperature decrease.

# 1. Introduction

Magnetic topological materials represent a burgeoning field of research that bridges the gap between topological quantum states and magnetism. These materials, characterized by their intrinsic magnetic orders and non-trivial topological electronic structures, have garnered significant attention for their potential to realize exotic quantum phenomena such as the quantum anomalous Hall effect, axion insulators, and Majorana fermions.[1–8] In europium (Eu) based magnetic topological materials, $Eu^{2+}$, with its strong magnetic moments ($4f^7$), plays a pivotal role in inducing and tuning the magnetic interactions within these compounds, thereby influencing their topological properties. The interaction between the localized magnetic moments of europium atoms and the itinerant electrons from other atoms in the crystal lattice can lead to a variety of magnetic ground states, including ferromagnetic, antiferromagnetic, and more complex configurations.[9] Numerous Eu-based ternary compounds have been identified to display magnetic and topological characteristics, predominantly incorporating pnictogens. This category includes families, such as $EuMnSb_2$[10,11] and $EuCd_2As_2$[12–14], where the structured alternation of magnetic and nonmagnetic layers plays a pivotal role in modulating their electronic properties. For example, the observation of anomalous negative magnetoresistance in the topological semimetal $EuMg_2Bi_2$[15,16] and the topological insulator $EuIn_2As_2$ is attributed to the strength of ferromagnetic interactions manifested within A-type antiferromagnetic ordering. However, very little research has been done to explore the electronic and magnetic properties in novel ternary phases containing chalcogenides and $Eu^{2+}$. Within this realm of investigation, the $MnBi_2X_4$ (X = Se, Te) systems have attracted significant interest owing to emergent topological phases intimately associated with their layered antiferromagnetic ordering. [17–23] This prompts an intriguing line of inquiry regarding substituting $Mn^{2+}$ with $Eu^{2+}$, aiming to explore the potential emergence of novel phases. Such a substitution could potentially alter the magnetic and electronic landscapes of these compounds, providing fertile ground for the discovery of new quantum states influenced by the unique magnetic characteristics of $Eu^{2+}$.

On the other hand, many Eu materials have pressure-sensitive Eu valence.[24–26] Previous studies showed that the $Eu^{2+}$ in $EuSn_2P_2$ [27] and $EuCd_2As_2$[28] are very robust under high pressure, up to 40 GPa. Thus, the question arises: can pressure tune the valence states in the new europium-based bismuth selenide?

In this study, we report the synthesis of a novel europium bismuth chalcogenide, $Eu_4Bi_6Se_{13}$, which exhibits low-field metamagnetic behavior along the crystalline *b*-axis and retains metallic conductivity throughout the entire temperature regime. This compound represents the first instance within the chalcogenide family - predominantly characterized by orthorhombic space group symmetry[29–31] - to crystallize in a monoclinic unit cell. Moreover, the pressure study demonstrates the mixed valence states of europium occur at much lower pressure compared to other Eu-based pnictogens. This observation expands the structural diversity observed in europium bismuth chalcogenides and underscores the unique magnetic and electronic phenomena inherent to this new material.

## 2. Results and Discussion

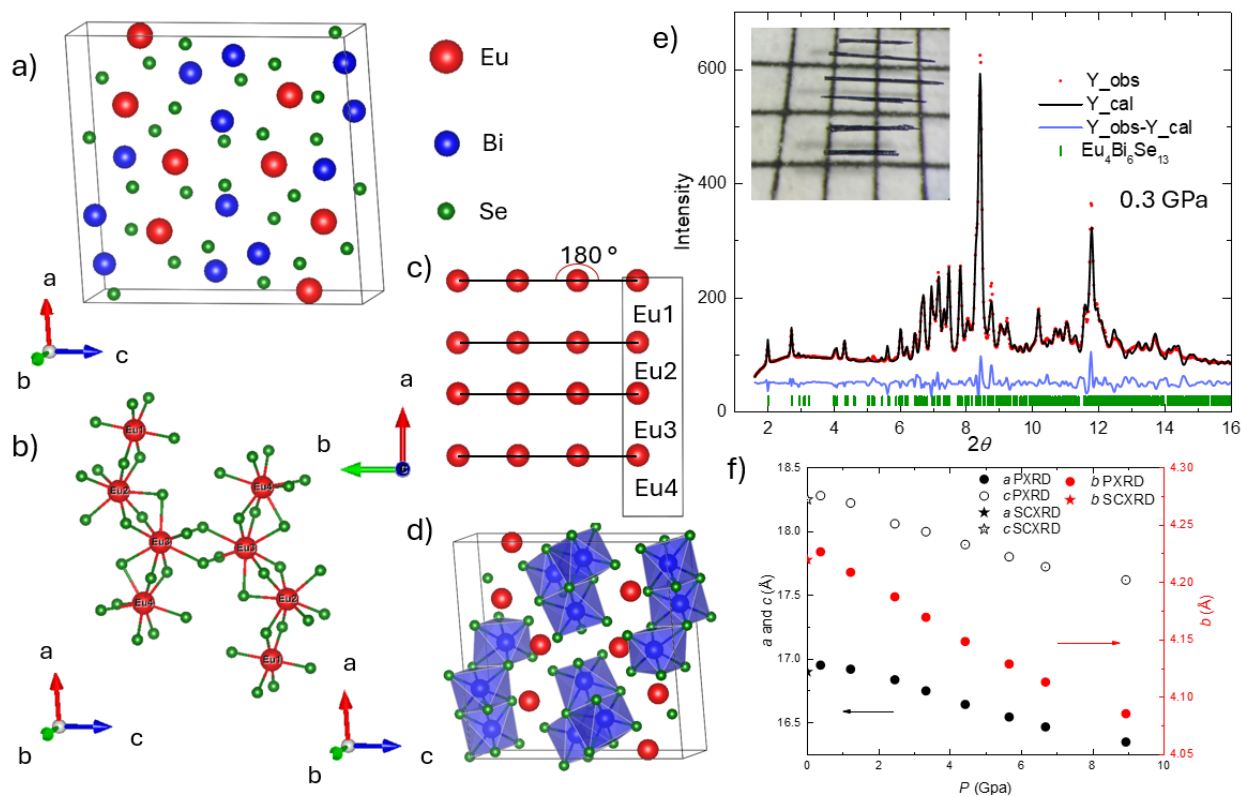

**Figure 1. Crystal structure of Eu$_4$Bi$_6$Se$_{13}$ and high-pressure powder X-ray diffraction measurements**. (The structure of Eu$_4$Bi$_6$Se$_{13}$ is shown in **Figure 1a**. The black line gives the unit cell. **Figure 1b** gives the coordination environment of each Eu site. The linear chains of Eu-Eu interactions extending along the *b*-axis are shown in **Figure 1c**. **Figure 1d** shows the edge-share distorted BiSe$_6$ octahedra. **Figure 1e** presents powder X-ray diffraction (PXRD) data at 0.3 GPa. The inset shows the picture of crystals on the millimeter grid paper. **Figure 1f** shows the lattice parameters comparison between single crystal X-ray diffraction (SCXRD) and PXRD under high pressure.)

Eu$_4$Bi$_6$Se$_{13}$ crystalized with the monoclinic *P*2$_1$/*m* space group, exhibiting isostructural properties with Sr$_4$Bi$_6$Se$_{13}$.[32] **Figure 1a** shows the crystal structure obtained from single crystal X-ray diffraction refinement (SCXRD data shown in **Table S1** and **S2**). As shown in **Figure 1b**, four different europium sites are surrounded by the bismuth sites. To the best of our understanding, this represents the inaugural instance of an Eu-Bi-Se adopting a monoclinic framework. Chemically analogous entities typically assume an orthorhombic architecture yet display congruent unit cell dimensions and structural motifs as identified in this novel configuration. As illustrated in **Figure 1b**, such motifs include relatively planar cells punctuated by Bi-Se connectivity, either in ladder or columnar arrangements, alongside extensive $Eu^{2+}$ coordination environments. Notably, the coordination geometries around $Eu^{2+}$ sites vary within the structure. The Eu1 atom is encased in a six-fold coordination by Se atoms, constituting a distorted octahedral geometry with an average bond length of 3.098(1) Å. In contrast, Eu2 and Eu4 atoms engage in an eight-fold coordination, forming a distorted square antiprism with a singular square face apparent. The Eu3 atom, uniquely, is nine-coordinated, adopting elongated bond lengths to maintain its divalent state, evidenced by an average bond length of 3.321(1) Å. As depicted in **Figure 1c**, Eu-Eu interactions yield linear chains extending along the *b*-axis, characterized by a consistent bond distance that aligns with the unit cell length of 4.219(1) Å. **Figure 1d** presents the edge-shared distorted $BiSe_6$ octahedra forming the octahedra chains and blocks. The Eu atoms are embedded in layers of the octahedra. Phase purity was ascertained through powder Rietveld refinement employing synchrotron data acquired from the Advanced Photon Source (APS) at Argonne National Laboratory (ANL), as illustrated in **Figure 1e**. The needle-shaped crystals exhibited feeble reflections in laboratory single-crystal and X-ray powder diffraction analyses, necessitating the application of higher energy and high flux synchrotron radiation. **Figure 1f** gives the *a*, *b*, and *c* lattice parameters under pressure compared with ambient pressure SCXRD data summarized in **Tables S1** and **S2**. Phase composition was further validated through SEM-EDS analysis, as depicted in **Figure S1**.

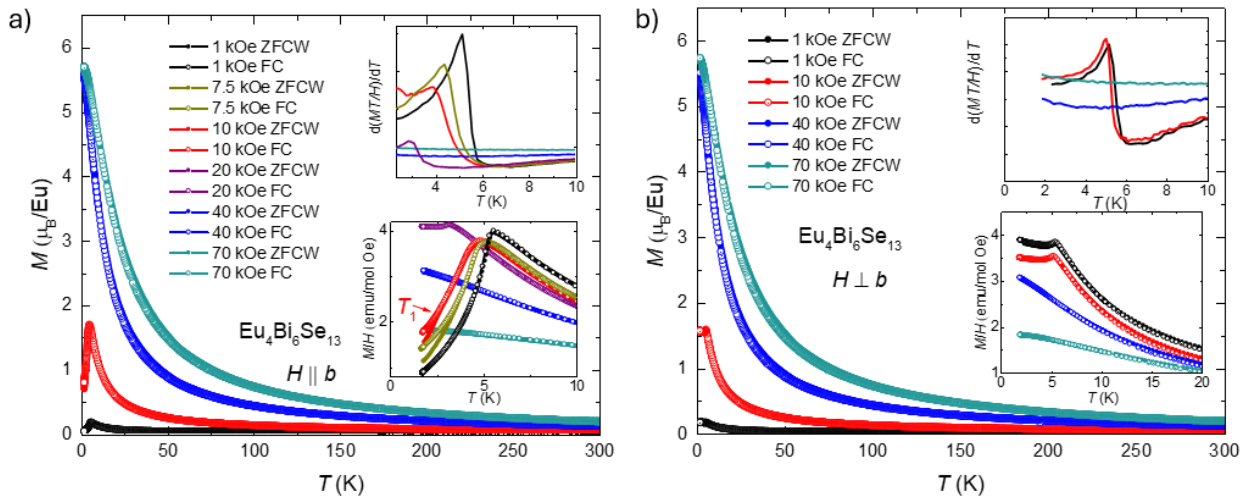

**Figure 2. Temperature-dependent magnetization in the different magnetic fields**. (Temperature-dependent magnetization in the different magnetic fields parallel (**Figure 2a**) or perpendicular (**Figure 2b**) to the *b*-axis are shown with Zero-Field-Cool-Warming (ZFCW) and

Field-Cool (FC) temperature protocols. The lower insets show the magnetization at a low-temperature range. The upper insets present the temperature derivative of the value $M \times T/H$.[33] ($M$ presents magnetization, $T$ presents temperature and $H$ presents magnetic field) $T_1$ gives the temperature at which the ZFCW and FC magnetization split. The criterion of $T_1$ is shown in the inset of **Figure S2a**.)

**Figures 2a** and **2b** present the temperature-dependent magnetization as the magnetic field parallel or perpendicular to the *b*-axis, respectively, with Zero-Field-Cool-Warming (ZFCW) and Field-Cool (FC) temperature protocols. There is no observable difference between ZFCW and FC measurements at temperatures from 5 K to 300 K with magnetic fields from 1 kOe to 70 kOe. All the magnetization shows the tail-like behavior as a function of temperature before a feature shown in the low magnetic field around 5 K. This feature shows a sudden decrease in magnetization, indicating the antiferromagnetic transition. When the magnetic field increased up to 40 kOe, no feature is observed above 1.8 K. Compared to the feature as the field applied perpendicular to the *b*-axis, the magnetization with the field parallel to the *b*-axis decreases by a much larger value, indicating a clear anisotropy. We assume this feature indicates the phase transition based on further investigation. If we define the transition temperature using the average onset and offset values, as shown in **Figure S2a**. The upper insets show a suppression of the transition temperature as the magnetic field increases. The transition temperature in the field parallel to the *b*-axis suppresses much faster than perpendicular to the *b*-axis. The lower insets show details of temperature-dependant magnetization at the low-temperature range. Except for a clear difference in the decreased value of magnetization after transition, the hysteresis is shown at a low temperature in the range of magnetic field of 1 kOe < H < 20 kOe with magnetic field direction parallel to the *b*-axis; However, in another direction, no hysteresis is observed. The discussion above suggests that the magnetic easy axis is along the *b*-axis.

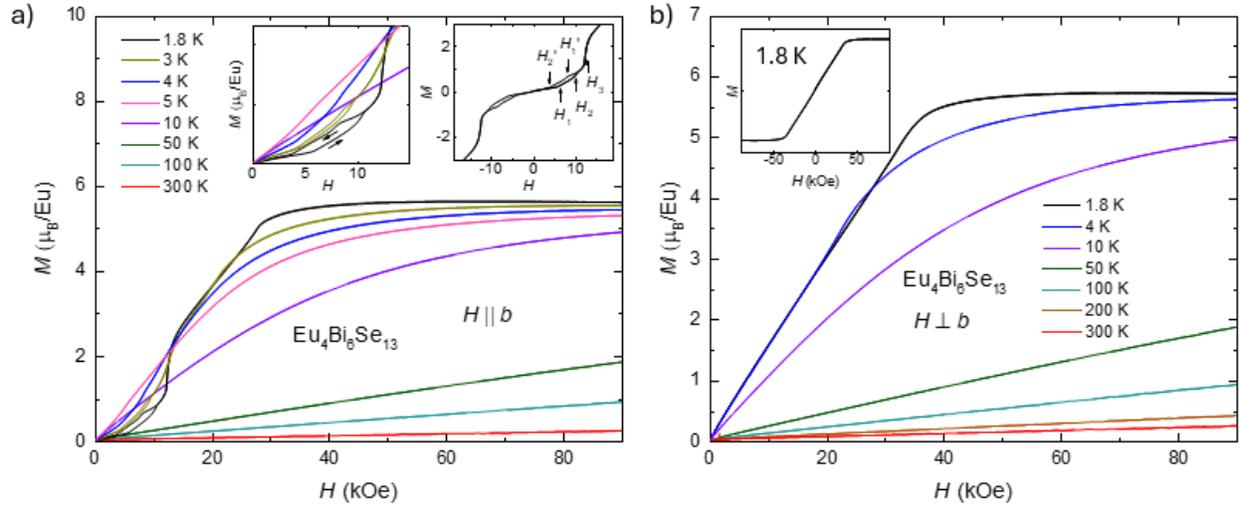

**Figure 3. Field-dependent magnetization at different temperatures**. (Magnetic field-dependent magnetizations in the different magnetic fields parallel (**Figure 3a**) or perpendicular (**Figure 3b**) to the *b*-axis are shown. The left inset of **Figure 3a** gives the details of low-field magnetization. The right inset in **Figure 3a** and the inset in **Figure 3b** show the full loop measurement of magnetization as a function of the magnetic field at 1.8 K with a certain field range (Measurements were taken up to 90 kOe.). The arrows in the right inset of **Figure 3a** indicate the direction of magnetic fields changing. The arrows give $H_1$, $H_2$, $H_1'$, $H_2'$, and $H_3$, which present the feature fields determined by **Figure 2S*b***.)

**Figure 3** presents the field-dependent magnetization at the different temperatures with the magnetic field parallel (**Figure 3a**) or perpendicular (**Figure 3a**) to the *b*-axis. As the magnetic field increases to 90 kOe, the magnetization is saturated when the temperature is below 5 K. As the temperature is above 50 K, linear field-dependent magnetization is observed in the field range of 0 kOe – 90 kOe. As the field is perpendicular to the *b*-axis, no hysteresis is shown. When the magnetic field is parallel to the *b*-axis direction, hysteresis is observed at a certain magnetic field and temperature range. At 1.8 K, no hypothesis is shown at a small field range below 3 kOe. When the field increases, the hysteresis loop shows. As the field continues to increase, the metamagnetic-like jump of magnetization is around 12 kOe. When the magnetic fields increase further, magnetization increases in the same slope as in another direction before saturation. The right inset of **Figure 3a** gives the full loop of magnetization as a function of the magnetic field at 1.8 K with a field up to 90 kOe (to show the hysteresis clearly, the range of the magnetic field shown in the inset is from -18 kOe to 18 kOe.). The saturation moment is 5.68 $\mu_B$/Eu as the field parallel to the *b*-axis, which is very similar to the saturation moment of 5.72 $\mu_B$/Eu as the field perpendicular to the $b$-axis. The arrows show the feature fields $H_1$, $H_2$, $H_1'$, $H_2'$, and $H_3$, which are determined by **Figure S2b**. As the temperature increases, the hysteresis size becomes smaller. At 5 K, there is almost no hysteresis observed. When the temperature increases to 5 K, the $H_1$, $H_2$, $H_1'$, $H_2'$, and $H_3$ features disappear. The field perpendicular to the $b$-axis measurements shows no hysteresis or jump-like feature.

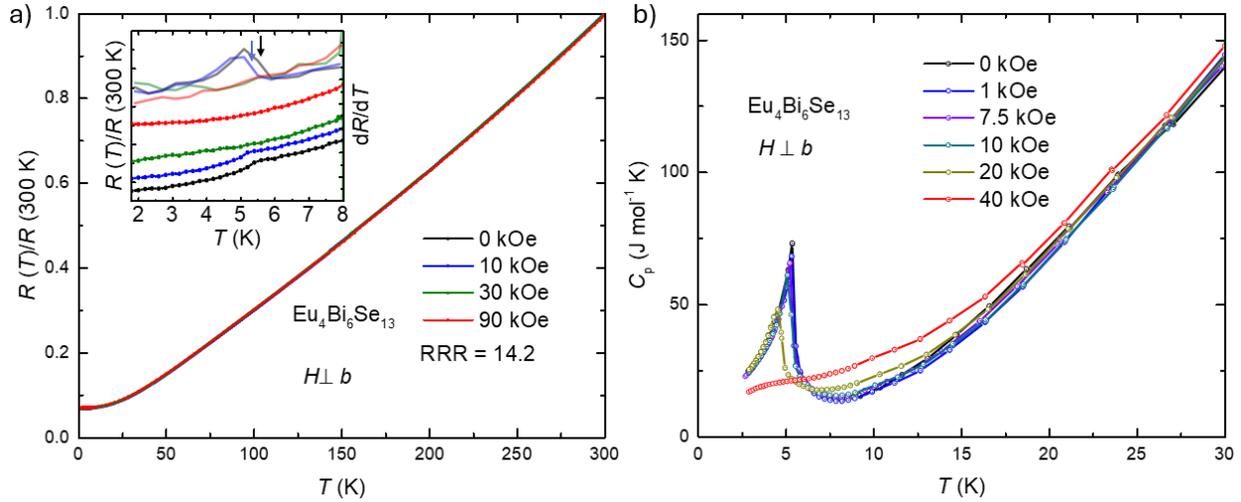

**Figure 4. Normalized resistance and specific heat measurements**. (Normalized resistance and specific heat measurements were taken in different magnetic fields perpendicular to the $b$-axis. **Figure 4a** shows the normalized resistance ($R(T)/R(300$ K$)$) as a function of temperature in the different magnetic fields perpendicular to the $b$-axis. The inset gives the low-temperature range of resistance (plotted with symbols and lines) and d$R$/d$T$ (plotted with transparency lines). The arrows indicate the transition temperatures. The color code of the inset is the same as **Figure 4a**. **Figure 4b** shows temperature-dependent specific heat in different magnetic fields perpendicular to the $b$-axis.)

**Figure 4a** presents the temperature-dependent resistance measurements in different magnetic fields perpendicular to the $b$-axis. The resistance as a function of temperature shows a metallic behavior with the RRR (residual resistance ratio) is 14.2. Overall, in the direction of $H \perp b$, the resistance does not change much under different fields; however, as shown in the inset, the kick-like features are observed in low-temperature and low-field, which are suppressed by a magnetic field. These features are clearer in the d$R$/d$T$ plot. These feature temperatures are around 5 K and decrease with magnetic field increase. The kink-like feature disappeared when we applied a 30 kOe magnetic field. **Figure 4b** shows temperature-dependent specific heat in the magnetic fields perpendicular to the *b*-axis. The second-order-like phase transition is observed in the field under 40 kOe. As the field increases, transition temperature decreases.

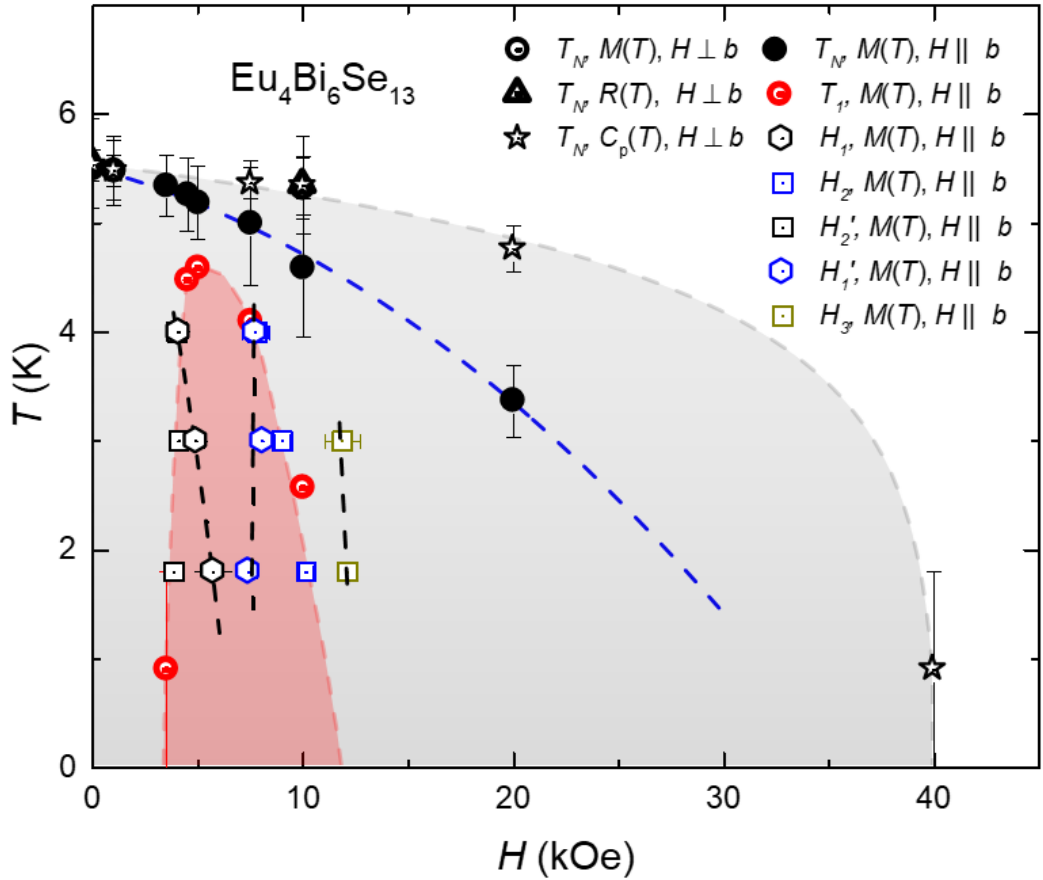

**Figure 5. Magnetic phase diagram of $Eu_4Bi_6Se_{13}$.** (The phase diagram is plotted based on the thermodynamic and transport measurements taken on single crystals of $Eu_4Bi_6Se_{13}$. The criteria of transition temperatures are shown in **Figure S2a**, **Figure S2b**, **Figure S3a**, and **Figure S3b**. The average of onset and offset values is transition temperature and the transition width is determined by the half of difference of onset and offset values. The transition temperature based on temperature-dependant magnetization using $d(M \times T/H)dT$ is given by round symbols. $H_1$, $H_2$, $H_1'$, $H_2'$, and $H_3$ are given by field-dependant magnetization measurements using $dM/dH$ in the direction of $H \parallel b$. The hollow triangle symbols show the transition temperatures determined by $dR/dT$ of temperature-dependent resistance measurement in the direction of $H \perp b$. The hollow star symbols present the transition temperatures based on temperature-dependent specific heat measurements. The black dashed line indicates the metamagnetic transition and the blue dashed line gives the transition in the $H \parallel b$ direction.)

**Figure 5** presents the phase diagram of Eu$_4$Bi$_6$Se$_{13}$ based on the thermodynamic and transport measurements. As the magnetic field is applied perpendicular to the *b*-axis, temperature-dependent magnetization, resistance, and specific heat measurements give similar phase transitions. The transition temperatures determined by d($M{\times}T/H$)d$T$, d$R$/d$T$, and $C_\mathrm{p}(T)$ overlap well in the phase diagrams. Considering the decrease of the magnetization, kink-like feature in resistance, and second-order-like transition in specific heat, we suspect this transition is antiferromagnetic. The Neel temperatures, $T_\mathrm{N}$, as the field perpendicular to the *b*-axis, are suppressed as the magnetic field increases. In the field parallel to the *b*-axis, the $T_\mathrm{N}$ differs from $H \perp b$. The decrease of $T_\mathrm{N}$ is faster as a function of the magnetic field compared with $H \parallel b$. Field-dependent magnetization has more features in the $H \parallel b$ direction than in another direction. As shown in **Figure 3a** and **Figure S2b**, when the measurements were taken as the field increased, a slope change happened at $H_1$. After that, a kink-like feature is shown at $H_2$, corresponding to the end of hysteresis. Then, a metamagnetic feature is shown at $H_3$ before the saturation. When the measurements are taken as the field decreases, the metamagnetic feature at $H_3$ does not change, but the magnetic fields of the other two features do change. As shown in **Figure S2b**, we use $H_1'$ and $H_2'$ to represent these features. **Figure 5** shows that the $T_1$ from the temperature-dependent magnetization well overlaps the $H_2$ and $H_2'$ tendencies, shown as the boundary of the red shadow. Since $T_1$ presents the hysteresis, this may indicate that $H_2$, $H_2'$, and $T_1$ correspond to the same physics phenomena: the existence of the magnetic domains and spin-flop. We suspect the appearance of magnetic domains is due to the competition among demagnetization, exchange, and magnetocrystalline energy. When the magnetic field increases, a net moment appears in the antiferromagnetic background, which may be due to spin-flop and gives the demagnetization energy. The magnetic domain appears if the demagnetization energy exceeds the domain wall's creation energy. $H_1$ and $H_1'$ should correspond to the motion of the domain wall since the tendency is the same with $H_2$ and $H_2'$. However, the magnetic domains are not shown in the field perpendicular to the *b*-axis. This may be due to the magnetocrystalline energy, making the extra energy cost to form the domain along the field direction. $H_3$ presents the metamagnetic transition, which does not show any hysteresis during the increase or decrease of the magnetic field.

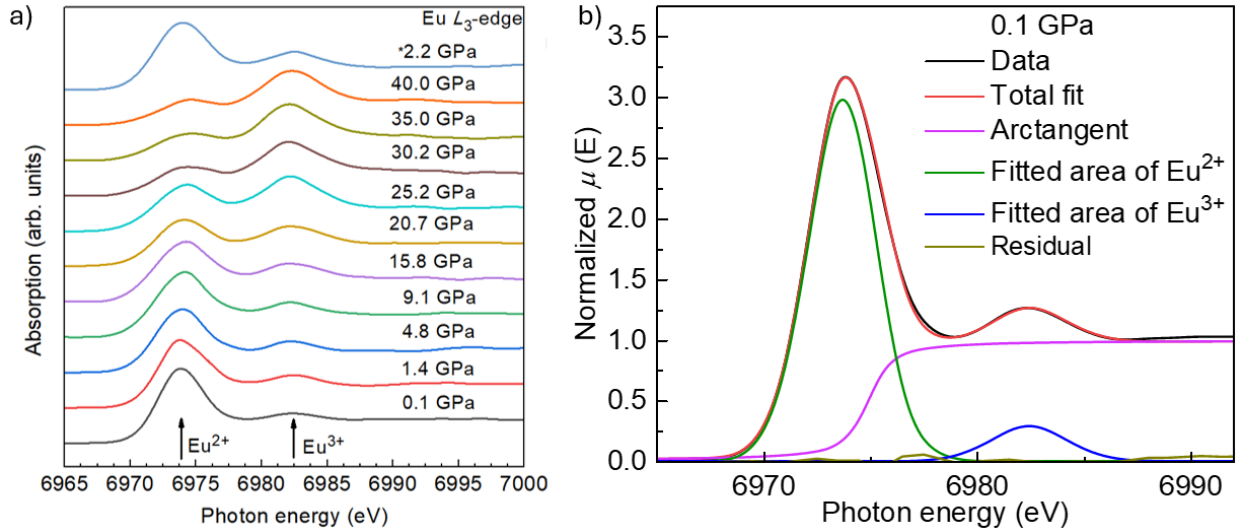

**Figure 6. High-pressure Partial Fluorescence Yield X-ray spectrometry (PYF-XAS) spectra and analysis**. **Figure 6a** shows the PYF-XAS spectra at the $L_3$ edge in $Eu_4Bi_6Se_{13}$ with increasing pressure from 0.1 GPa to 40.0 GPa along with a decompressed pressure of 2.2 GPa (marked with an asterisk). **Figure 6b** presents the analysis of the PFY-XAS data at 0.1 GPa using two sets of Lorentzian and arctangent functions for $Eu^{2+}$ and $Eu^{3+}$.

**Figure 6a** gives the PYF-XAS spectra of the $Eu_4Bi_6Se_{13}$ under pressure. Data was taken first with increasing pressure from 0.1 GPa to 40.0 GPa, then under decompression to 2.2 GPa. At 0.1 GPa, the Eu ions are mainly in the divalent state and with about 10 % of the trivalent state, which is evidenced by a dominant peak above 6.974 keV (corresponds to $2p^6_{3/2}4f^75d^06s^2 \rightarrow 2p^5_{3/2}4f^75d^16s^2$ transition) and a small absorption peak that appears at ~8 eV higher (corresponds to $2p^6_{3/2}4f^65d^16s^2 \rightarrow 2p^5_{3/2}4f^65d^26s^2$ transition). [34] As the pressure increases, the intensity of the $Eu^{2+}$ absorption peak decreases, and the intensity of the $Eu^{3+}$ absorption peak increases. This indicates a transition towards the trivalent state from the divalent state. When pressure is released, the divalent state of Eu ions is recovered, showing a reversible valence transition. The mean valence of Eu is derived by modeling the PFY-XAS data with one set of Lorentzian and arctangent functions for each peak, as shown in **Figure 6b**.

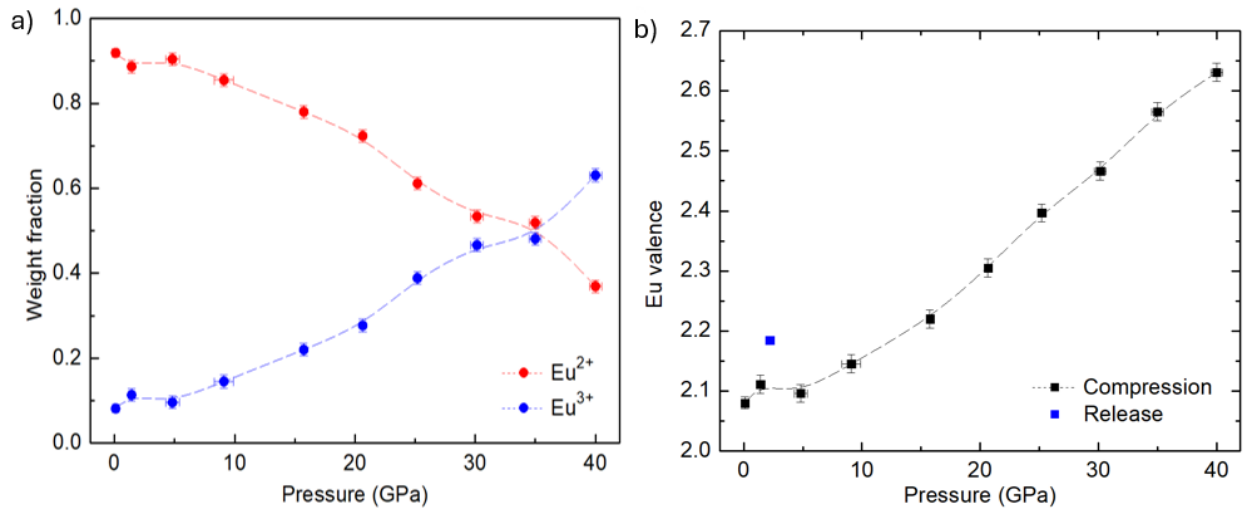

**Figure 7. Weight fraction and Eu valence as a function of pressure from PYF-XAS spectra analysis**. (Pressure-dependent weight fraction (**Figure 7a**) and Eu valence (**Figure 7b**) are shown using PYF-XAS spectra analysis. The blue point presents the average valence from releasing pressure.)

The weight fraction of $Eu^{2+}$ and $Eu^{3+}$ is given as a function of pressure using PYF-XAS spectra in **Figure 7a**. As the pressure increases, the fraction of $Eu^{2+}$ decreases with increasing of $Eu^{3+}$. At 35 GPa, the ratio of $Eu^{2+}$ and $Eu^{3+}$ is around 50%, and as the pressure is increased further, the ratio of $Eu^{3+}$ continues increasing. **Figure 7b** presents the Eu valence based on the fraction of $Eu^{2+}$ and $Eu^{3+}$. The black points give the pressure-increasing measurements, and the blue point presents the pressure-decreasing measurement. The valence after pressure release goes back to the $Eu^{2+}$ rich, indicating the reversible valence transition.

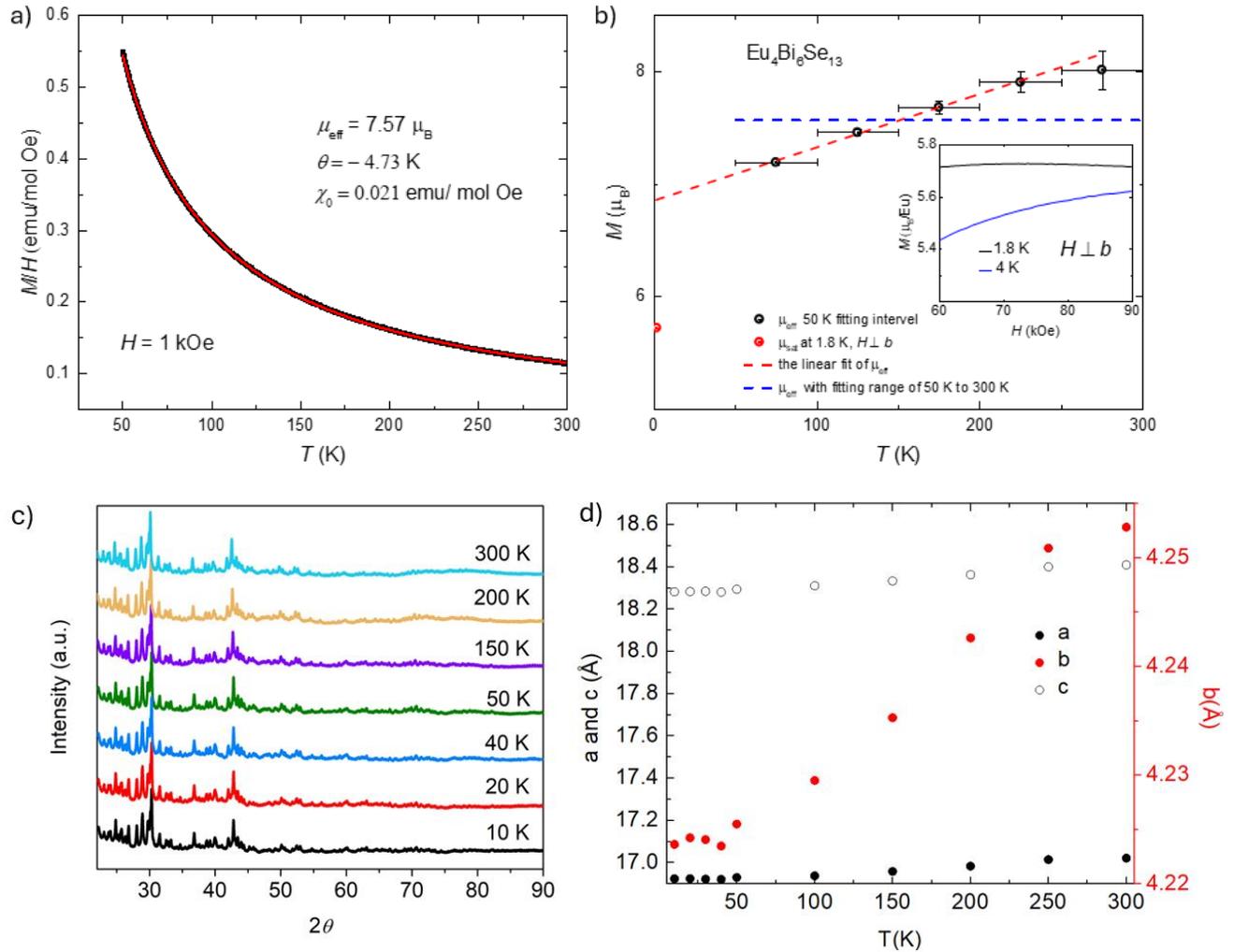

**Figure 8. Curie-Weiss (CW) fitting on the polycrystalline average susceptibility (Figure 8*a*), the effective moments compared with saturation moments (Figure 8*b*), and temperature-dependent PXRD measurements. Figure 8*a*** presents the CW fitting with the range of temperature 50 K - 300 K on a polycrystalline average susceptibility calculated as $\chi = (2\chi_\perp + \chi_\parallel)/3$, where $\chi_\perp$ and $\chi_\parallel$ denote susceptibilities perpendicular and parallel to the crystallographic *b*-axis, respectively. **Figure 8*b*** compares the effective moments from different CW fitting temperature ranges and saturation moment at 1.8 K. Inset shows the field-dependent magnetization. **Figure** 8*c* shows the temperature-dependent PXRD patterns, and **Figure 8*d*** presents the lattice parameters change given by GSAS II fitting.

**Figure 8*a*** shows the Curie-Weiss (CW) fitting with a temperature range of 50 K – 300 K, and CW analysis is on the polycrystalline average susceptibility with 1 kOe magnetic field calculated as $\chi = (2\chi_\perp + \chi_\parallel)/3$ (β is close to 90 degrees and the *a* and *c* lattice parameters are similar), where $\chi_\perp$ and $\chi_\parallel$ denote susceptibilities perpendicular and parallel to the crystallographic *b*-

axis, respectively. The $\mu_{eff}$ is 7.57 $\mu_B$, slightly less but close to the theoretical values of 7.94 $\mu_B$ value of $Eu^{2+}$.[35] If we assume that $Eu^{3+}$ gives zero effective moment (J = 0), 7.57 $\mu_B$ gives around 95% $Eu^{2+}$ and 5% $Eu^{3+}$. Considering that this system may not be accurately described by Hund's law due to the crystal field, we use the ambient pressure PYF-XAS data to calibrate the effective moment. We fit the susceptibility from 50 K to 300 K with 50 K intervals, as shown in **Figure 8b**. Then, we use the fitting value at the temperature range of 250 K -300 K to calibrate the $Eu^{2+}$ effective moment, which is 8.71 $\mu_B$. Comparing this with the saturation moment at 1.8 K with the field perpendicular to the *b*-axis (saturation moments of $H \parallel b$ and $H \perp b$ are similar, and field-dependent magnetization at $H \perp b$ doesn't have domain motion.), as shown in **Figure 8b**, the effective moment of high temperature is larger than the saturation moment at 1.8 K, which is only 5.72 $\mu_B$. Using the saturation moment and calibrated $Eu^{2+}$ effective moment, the ratio of $Eu^{2+}$ at 1.8 K is only 66.5%. If we assume the dominant determination of change in valence under high-pressure and at room temperature is the lattice contraction, temperature-dependent valence change at low temperature should have a different mechanism, since, according to **Figure 1f**, **Figure 8c**, and **Figure 8d**, the lattice parameters impact by temperature is much smaller than pressure (up to 10 GPa), though the impact on lattice parameters from decreasing the temperature decreases and increasing pressure have the same trend. Especially, at low temperatures, there is no observed lattice parameter change below 50 K. However, the large valence change may appear according to the CW fitting and saturation moments. As shown in **Figure 3a** and the inset of **Figure 8b**, a clear decrease of saturation moment when temperature increases in the order state indicates the opposite temperature-dependant valence change in higher temperature given by CW fitting. All of these suggested that, in the low-temperature range, the valence of Eu is not dominantly affected by the lattice parameter changes as shown by high-pressure measurements but influenced by the electron structure, which is the hybridization of conduction election and 4*f* electrons.

## 3. Conclusion

In this study, we successfully synthesized Eu$_4$Bi$_6$Se$_{13}$, a novel metallic material, using the solid-state reaction method. This compound exhibits a monoclinic crystal structure within the *P*2$_1$/*m* space group, distinguishing it from other europium bismuth chalcogenides in structural uniqueness. Notably, the material transitions into an antiferromagnetic state near 5 K and demonstrates magnetic anisotropy. The magnetic easy axis is aligned with the monoclinic axis. Magnetic characterization at low applied magnetic fields revealed metamagnetic transitions attributable to spin reorientation phenomena, likely due to the material's inherent weak antiferromagnetic exchange relative to magnetic anisotropy. Phase-diagram is plotted with a magnetic domain motion that appears under a certain magnetic field and temperature ranges. PFY-XAS studies showcased a monotonic valency change towards a trivalent state with increasing pressure. This phenomenon is considered to be a result of the volume collapse of an intermediate valence state. The electronic structure change at the low temperature induces the temperature-dependent valence change, which is shown in the CW fitting of the temperature-dependent magnetic susceptibility data. This, which is different from pressure-dependent valence change, is further proved by temperature-dependent PXRD measurements.

## 4. Methods

**4.1. Crystal Growth:** Needle-shaped crystals of $Eu_4Bi_6Se_{13}$ were synthesized utilizing a solid-state reaction technique. An ingot of Europium (sublimed dendritic, REO grade, sourced from Thermo Scientific), finely ground Bismuth pieces (purity of 99.99%, provided by Strem Chemicals Inc.), and Selenium powder (with a purity of 99.995%, obtained from Thermo Scientific) with stoichiometry ratio were homogeneously mixed and pressed into pellets, each weighing about 1000 mg. These pellets were put in an alumina crucible and sealed in evacuated quartz tubes (with a vacuum level maintained below $10^{-5}$ torr) to prevent oxidation and contamination. The ampoule was heated to 800 °C at a ramp rate of 60°C per hour, maintained at 800 °C for a duration of 48 hours to ensure a complete reaction, and then cooled down at the same rate to room temperature. The needle-shaped single-crystalline $Eu_4Bi_6Se_{13}$ was obtained with an extra EuSe phase. Since the shape is quite different, $Eu_4Bi_6Se_{13}$ can be mechanically separated from the EuSe phase. The separated pure $Eu_4Bi_6Se_{13}$ crystals are used in all the measurements.

**4.2. Structural and Chemical Composition Determination:** The single crystal X-ray diffraction of $Eu_4Bi_6Se_{13}$ was taken using a Rigaku XtaLab Synergy-S X-ray diffractometer equipped with Mo radiation ($\lambda_{K\alpha}$ = 0.71073 Å) and an Oxford Cryosystems 800 low-temperature device. The selected pure-phase single crystal was mounted on a nylon loop with PARATONE oil. To minimize data acquisition time and prevent the excessive accumulation of peaks arising from the low-symmetry monoclinic cell, measurements were carried out at 100 K, with an exposure duration of 5.0 seconds per frame. The total number of runs and images was were based on the strategy calculation from the program CrysAlisPro 1.171.43.92a (Rigaku OD, 2023). Data reduction was performed with correction for Lorentz polarization. A numerical absorption correction was applied based on Gaussian integration over a multifaceted crystal model.[36] Empirical absorption correction used spherical harmonics, implemented in the SCALE3 ABSPACK scaling algorithm.[37] The structure was solved and refined using the SHELXTL Software Package.[38,39]

The sample phase composition was analyzed employing a JEOL 6610LV scanning electron microscope equipped with a tungsten hairpin emitter (JEOL Ltd., Tokyo, Japan). For elemental analysis, energy-dispersive X-ray spectroscopy (EDX) was conducted utilizing an Oxford Instruments AZtec system (Oxford Instruments, High Wycomb, Buckinghamshire, England), operating software version 3.1. This setup included a 20 mm² Silicon Drift Detector (SDD) and an ultra-thin window integrated with the JEOL 6610LV SEM. The needle-like crystals were affixed to carbon adhesive tape and introduced into the SEM chamber for examination at an accelerating voltage of 20 kV. Data acquisition entailed collecting spectra at multiple points along the individual crystals over an optimized timeframe. Quantitative compositional analysis was performed using SEM Quant software, which applies corrections for matrix effects to the intensity measurements.

**4.3. Phase Identification:** Given a small amount of needle-like crystals and the significant background interference from laboratory Cu Kα radiation, high-pressure phase analysis uses synchrotron powder X-ray diffraction. This was obtained from pulverized needle-like crystals at the 13BM-C beamline (PX2) of the Advanced Photon Source (APS) at Argonne National Laboratory (ANL), employing a wavelength of 0.434 Å. A BX-90 Diamond Anvil Cell, equipped with a 200 μm culet, was utilized to apply pressure. A 4:1 (volume ratio) methanol-ethanol mixture was used as the pressure-transmitting medium. The two-dimensional diffraction images were integrated using Dioptas software, and Rietveld refinement of the dataset was conducted with GSAS II.[40]

The temperature-dependent PXRD patterns were measured using a HUBER X-ray diffractometer with Cu Kα radiation ($\lambda = 1.5460$Å), equipped with a helium cryogenic system. A step size of 0.005 ° was used to measure spectra over a Bragg angle (2θ) range of 4-100 °. The powdered sample was measured every 10 K from 300 K to 10 K. The powder data was refined through the Rietveld method using the GSASII software.

**4.4. Magnetic and Electronic Properties Measurements:** The temperature- and field-dependent VSM magnetization measurements and resistance measurements, as well as the temperature-dependent specific heat of $Eu_4Bi_6Se_{13}$ single crystals, were carried out using a Quantum Design, Physical Property Measurement System (PPMS-DynaCool). In temperature- and field-dependent magnetization, the needle-shaped single crystal samples were arranged carefully in parallel orientation on Kapton Tape and secured on a quartz paddle sample holder. The magnetic signal from the tape is not considered. Around 1 mg samples were measured parallel and perpendicular to the b-axis with a magnetic field up to 90 kOe. DC electrical resistance measurements were performed in a standard four-contact geometry with a 1 mA current. 50 µm diameter Pt wires were bonded to the samples with silver conductive epoxy covered by silver paint (DuPont 4929N) with contact resistance values of about 2-3 Ohms. Up to 90 kOe, the magnetic field was applied perpendicular to the *b*-axis, with the current flowing parallel to the *b*-axis. Temperature-dependent specific heat measurements were carried out using the relaxation technique as implemented in the Heat Capacity option of the PPMS.

**4.5. High-Pressure Valence States Determination:** To provide direct information on the valence of Eu in the compound under pressure, the PFY-XAS (Partial fluorescence Yield X-Ray Spectrometry) experiment at the $L_3$ edge (6.974 keV) was conducted at room temperature at Beamline 16ID-D of the APS, ANL. The beam size was focused on a 4×6 µm area. The incident X-ray energy was scanned from 6.956 keV to 7.032 keV. To avoid strong absorption by the diamond anvils, incoming X-rays traveled through a Be gasket, and the fluorescence signal was collected. The signal was first analyzed with a silicon analyzer and then collected with a Pilatus detector at 90 degrees from the incident X-ray beam. Since the measurements were conducted in fluorescence mode, attenuation of fluorescence may occur as it travels out of the sample. The optimal sample position was determined by scanning the sample position to minimize this attenuation. High pressure was achieved using a symmetric-type diamond anvil cell equipped with a pair of diamond anvils with 300 µm culet. A beryllium gasket was pre-indented to ∼50 µm, and a hole of 200 µm was drilled. Cubic boron nitride (cBN) and epoxy mixture were then filled in the gasket hole and indented to form the gasket insert and maintain a good sample thickness under pressure. A 40 × 70 µm $Eu_4Bi_6Se_{13}$ needle was loaded in the sample chamber formed by the cBN insert along with a ruby ball used as an in-situ pressure marker. No pressure medium was used. Data was taken under compression up to 40 GPa and then under decompression to 2.2 GPa, with one point taken at the lowest pressure under decompression. Data was normalized in Athena, a part of the Demeter software package for XAS data processing and analysis.[41] Eu mean valence was obtained by modeling the XAS spectrum with two sets of Lorentzian and arctangent functions for divalent and trivalent states.

## Acknowledgment

Mingyu Xu, Jose L. Gonzalez Jimenez contributed equally to this work. The work at Michigan State University was supported by the U.S.DOE-BES under Contract DE-SC0023648. G.J., and W.B. were supported by the National Science Foundation (NSF) CAREER Award No. DMR-2045760. C. H. and X. L. were supported by the NSF PREC program No. 2216807. The work at University of Tennessee was supported by U.S. Department of Energy (DOE), Office of Science, Office of Basic Energy Sciences (BES), Materials Sciences and Engineering Division under Contract DE-SC0020254. X.K. acknowledges the financial support by the U.S. DOE-BES under DE- SC0019259. M.L. acknowledges support from DOE BES Award No. DE-SC0020148. This research used resources of the Advanced Photon Source (APS), a U.S. Department of Energy (DOE) Office of Science User Facility operated for the U.S. DOE Office of Science by Argonne National Laboratory (ANL) under Contract No. DE-AC02-06CH11357. Portions of this work were performed at HPCAT (Sector 16), APS, ANL. HPCAT operations are supported by DOE-NNSA's Office of Experimental Sciences. Support from COMPRES under NSF Cooperative Agreement No. EAR-1606856 is acknowledged for the PX$^2$ program. We thank Dongzhou Zhang and Yuming Xiao for assistance in synchrotron X-ray experiments.